\documentclass{PoS}
\usepackage{amsmath}
\usepackage{graphicx}
\usepackage{epsfig}

\title{Precise Charm- and Bottom-Quark Masses:\\ Recent Updates}

\ShortTitle{Precise Charm- and Bottom-Quark Masses: Recent Updates}

\author{\speaker{J. H. K\"uhn}\\
       Karlsruhe Institute of Technology}

\abstract{
Recent theoretical and experimental improvements in the determination of
charm and bottom quark masses are discussed. The final results,
$m_c(3\,\text{GeV})=986(13)\,$MeV and $m_b(m_b)=4163(16)\,$MeV represent, together 
with a closely related lattice determination $m_c(3\;{\rm GeV})=986(10)\,$MeV,
the presently most precise determinations of these two fundamental Standard Model
parameters. A critical analysis of the theoretical and experimental uncertainties 
is presented.
}

\FullConference{ RADCOR 2009 - 9th International Symposium on Radiative Corrections 
                 (Applications of Quantum Field Theory to Phenomenology) ,\\
		 October 25 - 30 2009\\
		 Ascona, Switzerland}

\begin{document}

The past years have witnessed significant improvement in the
determination of charm and bottom quark masses as a consequence of
improvements in experimental techniques as well as theoretical
calculations. 
%
Quark mass determinations can be based on a variety of
observables and theoretical calculations. The one presently most precise 
follows an idea advocated by the ITEP group more than thirty years ago 
\cite{Shifman:1978bx}, and has gained renewed interest after significant
advances in higher order perturbative calculations 
\cite{Marquard}
have been achieved.  
In particular the four-loop results (i.e. the coefficients $C_n$ discussed 
below) are now available for the Taylor coefficients of the vacuum 
polarization, analytically up to $n=3$ and numerically up to $n=10$.
The method exploits the fact that the vacuum polarization 
function $\Pi(q^2)$ and its derivatives, evaluated at $q^2=0$, can be considered
short distance quantities with an inverse scale characterized by the
distance between the reference point $q^2=0$ and the location of the
threshold $q^2=(3\;{\rm GeV)}^2 $ and $q^2=(10\;{\rm GeV})^2$ for charm and
bottom, respectively. This idea has been taken up in \cite{Kuhn:2001dm} after the
first three-loop evaluation of the moments became available 
\cite{Chetyrkin:1995ii,Chetyrkin:1996cf,Chetyrkin:1997mb} and
has been further improved in \cite{Kuhn:2007vp} using four-loop results
\cite{Chetyrkin:2006xg,Boughezal:2006px} for the
lowest moment. An analysis which is based on the most recent theoretical
\cite{Maier:2008he,Maier:2009fz,Kiyo:2009gb} and experimental progress has 
been performed in \cite{Chetyrkin:2009fv}
and will be reviewed in the following.

Let us recall some basic notation and definitions. The vacuum
polarization $\Pi_Q(q^2)$ induced by a heavy quark $Q$ with charge
$Q_Q$ (ignoring in this short note the so-called singlet contributions),
is an analytic function with poles and a branch cut at and above
$q^2=M_{J/\psi}^2$. Its Taylor coefficients $\bar C_n$, defined through
\begin{equation}
  \label{eq:Pi}
  \Pi_Q(q^2)\equiv Q_Q^2\frac{3}{16\pi^2} \sum_{n\geq0}\bar{C}_nz^n
\end{equation}
can be evaluated in pQCD, presently up to order $\alpha_s^3$.
Here $z\equiv q^2/4m_Q^2$, where $m_Q=m_Q(\mu)$ is the running
$\overline{\text{MS}}$ mass at scale $\mu$. Using a once-subtracted dispersion
relation
\begin{equation}
  \label{eq:disp_rel}
  \Pi_Q(q^2)=\frac{1}{12\pi^2}\int^\infty_0 ds \frac{R_Q(s)}{s(s-q^2)}
\end{equation}
(with $R_Q$ denoting the familiar $R$-ratio for the production of heavy quarks 
with flavour $Q$), the Taylor coefficients can be expressed through moments of 
$R_Q$. Equating perturbatively calculated and experimentally measured moments,
\begin{equation}
  \label{eq:M_exp}
  {\cal M}_n^{\text{exp}}=\int \frac{ds}{s^{n+1}}R_Q(s)
\end{equation}
leads to an ($n$-dependent) determination of the quark mass
\begin{equation}
\label{eq:m_Q}
m_Q= 
\frac{1}{2}\left(
\frac{9Q_Q^2}{4}\frac{C_n}{{\cal M}_n^{\text{exp}}}
\right)^{\frac{1}{2n}}\,.
\end{equation}

Significant progress has been made in the perturbative evaluation of
the moments since the first analysis of the ITEP group. The 
${\cal O}(\alpha_s^2)$ contribution (three loops) has been evaluated more
than 13 years ago \cite{Chetyrkin:1995ii,Chetyrkin:1996cf,Chetyrkin:1997mb}, 
as far as the terms up to $n=8$ are
concerned, recently even up to $n=30$ \cite{Boughezal:2006uu,Maier:2007yn}.  
About ten years later the
lowest two moments ($n=0,1$) of the vector correlator were evaluated
in ${\cal O}(\alpha_s^3)$, i. e. in four-loop approximation
\cite{Chetyrkin:2006xg,Boughezal:2006px}. 
The corresponding two lowest moments for the pseudoscalar
correlator were obtained in \cite{Sturm:2008eb} in order to derive the charmed
quark mass from lattice simulations \cite{Allison:2008xk}.  
In \cite{Maier:2008he,Maier:2009fz} the second and
third moments were evaluated for vector, axial and pseudoscalar
correlators. Combining, finally, these results with information about
the threshold and high-energy behaviour in the form of a Pad\'e
approximation, the full $q^2$-dependence of all four correlators was
reconstructed and the next moments, from four up to ten, were obtained
with adequate accuracy~\cite{Kiyo:2009gb}.

%

Most of the experimental input had already been compiled and exploited
in \cite{Kuhn:2007vp}, where it is described in more detail. However, until recently
the only measurement of the cross section above but still close to the 
$B$-meson threshold was performed by the CLEO collaboration more than
twenty years ago \cite{Besson:1984bd}. Its large systematic uncertainty was 
responsible for a
sizable fraction of the final error on $m_b$.  This measurement has been
recently superseded by a measurement of BABAR \cite{:2008hx} with a
systematic error between 2 and 3\%. In \cite{Chetyrkin:2009fv} the radiative
corrections were unfolded and used to obtain a significantly improved
determination of the moments.
\begin{figure}
  \centering
  \includegraphics[width=\linewidth]{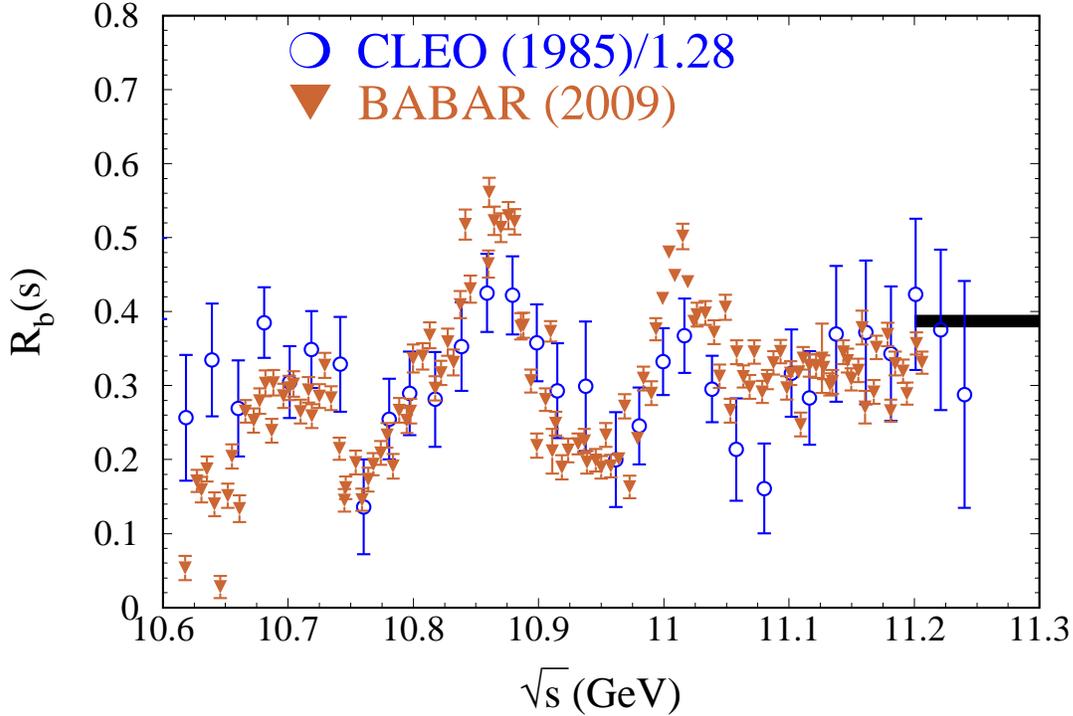}
  \caption{Comparison of rescaled CLEO data for $R_b$ with BABAR data.
    \cite{Chetyrkin:2009fv,:2008hx}. The black bar on the right corresponds to
    the theory prediction~\cite{Harlander:2002ur}.}
  \label{fig:R_b}
\end{figure}
\begin{table}
  \centering
    \begin{tabular}{l||r@{}l|rrrr|c||l|rrr|l}
\hline
$n$ & \multicolumn{2}{|c|}{$m_c(3~\text{GeV})$} & 
exp & $\alpha_s$ & $\mu$ & np & 
total &
$m_b(10~\text{GeV})$ & 
exp & $\alpha_s$ & $\mu$ &
total
\\
\hline
       1&   &986&  9&   9&   2&  1 & 13 &  3597&  14&  7&   2&  16\\
       2&   &976&  6&  14&   5&  0 & 16 &  3610&  10&  12&  3&  16\\
       3&   &978&  5&  15&   7&  2 & 17 &  3619&  8&  14&   6&  18\\
       4&  1&004&  3&   9&  31&  7 & 33 &  3631&  6&  15&  20&  26\\
\hline
\end{tabular}
\caption{Results for $m_c(3~\text{GeV})$ and for $m_b(10~\text{GeV})$ in MeV.
The errors are from experiment,  $\alpha_s$, the variation of $\mu$ and (for 
$m_c$) the gluon condensate.
}
\label{tab:mc1}
\end{table}
%
The final results for $m_c (3\;{\rm GeV})$ and $m_b (10\;{\rm GeV})$ are
listed in Table \ref{tab:mc1}. Despite the significant
differences in the composition of the errors, the results for different
values of $n$ are perfectly
consistent. For charm the result 
from $n=1$ has the smallest dependence on the strong coupling and the smallest
total error, which we take as our final value 
\begin{equation}
\label{eq:mc_3}
m_c(3~\text{GeV}) = 986(13)~\text{MeV}\,,
\end{equation}
and consider its  consistency with $n=2$, $3$ and $4$ as additional
confirmation. Transforming this to the scale-invariant mass 
$m_c(m_c)$~\cite{Chetyrkin:2000yt}, including the
four-loop coefficients  of the renormalization group functions one
finds \cite{Chetyrkin:2009fv}
$m_c(m_c)=1279(13)~\text{MeV}$. Let us recall at this point that a recent
study \cite{Allison:2008xk}, 
combining a lattice simulation for the data for the pseudoscalar correlator
with the perturbative  three- and four-loop result
\cite{Chetyrkin:1997mb,Sturm:2008eb,Maier:2009fz} has led to
$m_c(3~\text{GeV})= 986(10)$~MeV in remarkable
agreement with \cite{Kuhn:2007vp,Chetyrkin:2009fv}.  

The treatment of the bottom quark case proceeds along
similar lines. However, in order to suppress the theoretically evaluated
input above 11.2 GeV (which corresponds to roughly 60\% for the lowest,
40\% for the second and 26\% for the third moment), the result from the
second moment has been adopted as our final result,
\begin{equation}
m_b(10~\text{GeV}) =3610(16)~\text{MeV},
\label{eq:m_b}
\end{equation}
corresponding to $m_b(m_b)=4163(16){\rm MeV}$. The explicit $\alpha_s$
dependence of $m_c$ and $m_b$ can be found in
\cite{Chetyrkin:2009fv}. When considering the ratio of charm and bottom
quark masses, part of the  $\alpha_s$ and of the $\mu$ dependence cancels
\begin{equation}
  \label{eq:m_ratio}
  \frac{m_c(3~\text{GeV})}{m_b(10~\text{GeV})}=
  0.2732 -\frac{\alpha_s-0.1189}{0.002}\cdot 0.0014 \pm 0.0028 
  \,,
\end{equation}
which might be a useful input in ongoing analysis of bottom decays.

\vspace{-.5cm}
\begin{figure}[ht]
\begin{center}
\begin{tabular}{cc}
\hspace{-1.cm}
\leavevmode
\epsfxsize=0.56\textwidth
\epsffile{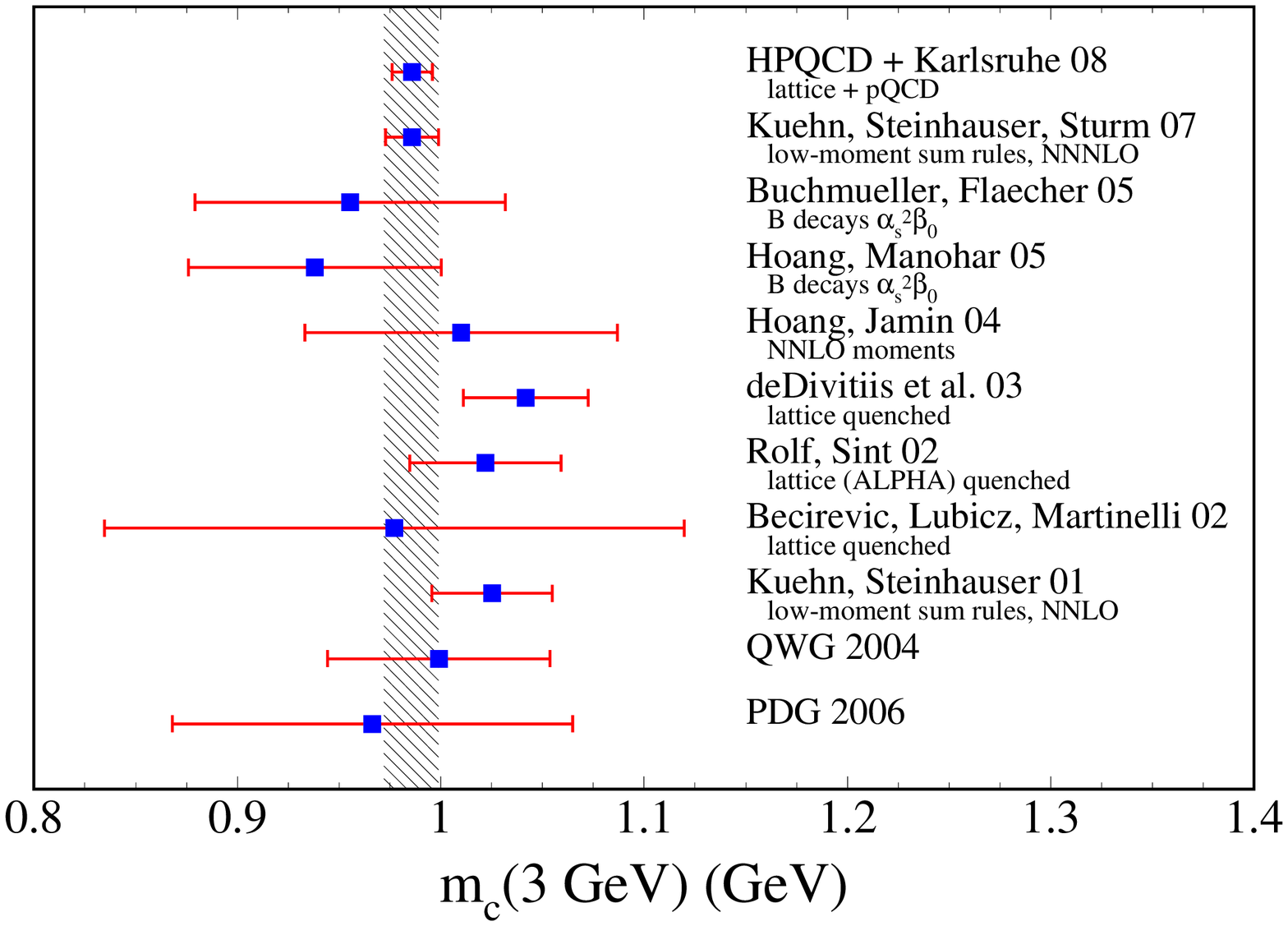}
&
\hspace{-1.cm}
\leavevmode
\epsfxsize=0.56\textwidth
\epsffile{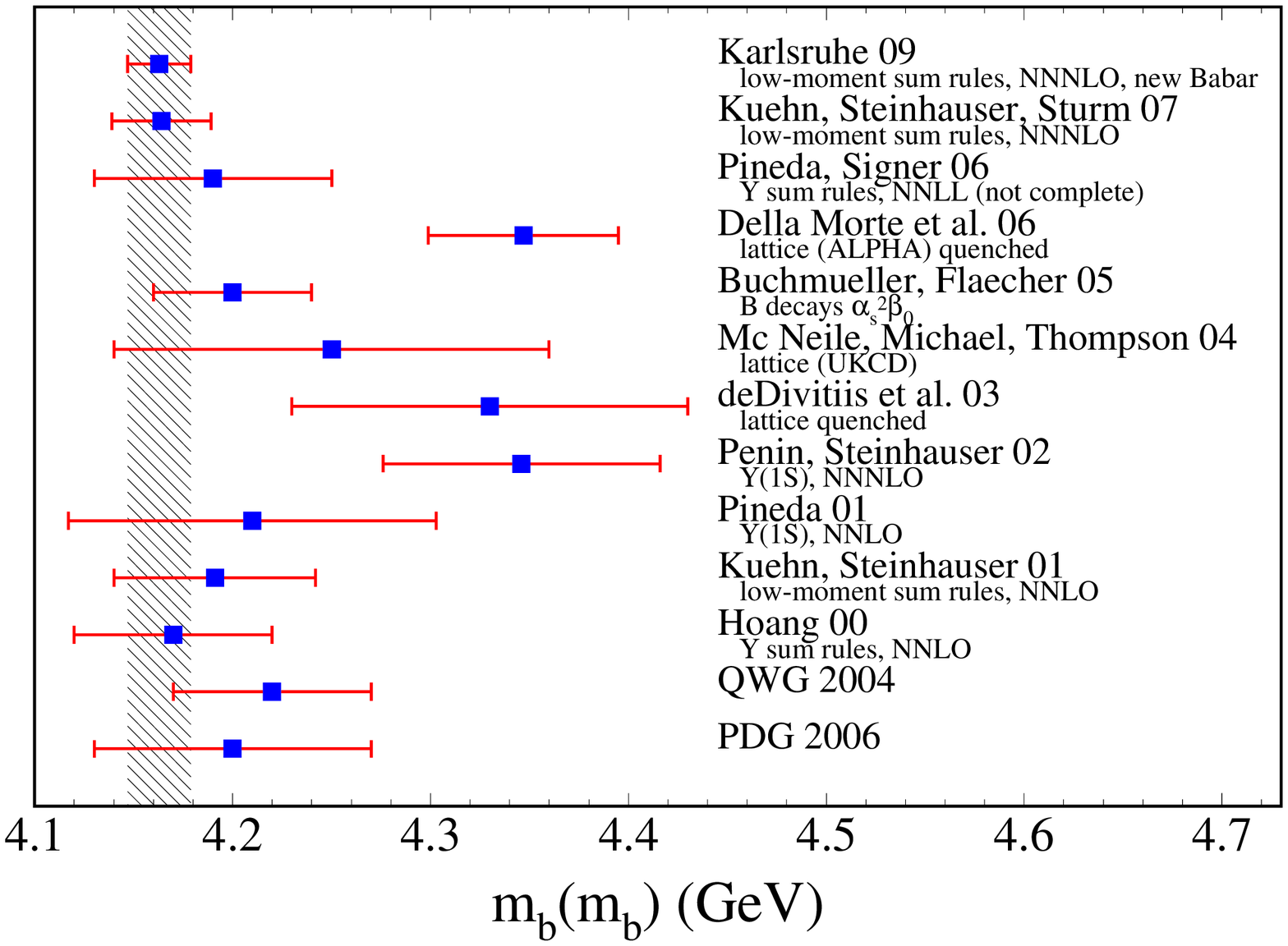}
\end{tabular}
\end{center}
\vspace*{-2em}
\caption{
Comparison of recent determinations of $m_c(3\;{\rm GeV})$ and $m_b(m_b)$.
}
\label{fig::mc_compare}
\end{figure}

In Fig.~\ref{fig::mc_compare} the results of this analysis are 
compared to others based on completely different methods. The $m_c$ value is well 
within the range suggested by other determinations. In case of $m_b$ our result 
is somewhat towards the low side, although still consistent with most other 
results.

The results presented in \cite{Chetyrkin:2009fv} constitute the most 
precise values for the
charm- and bottom-quark masses available to date. Nevertheless it is
tempting to point to the dominant errors and thus identify potential
improvements.  In the case of the charmed quark the error is dominated
by the parametric uncertainty in the strong coupling
$\alpha_s(M_Z)=0.1185\pm0.002$. Experimental and theoretical errors are
comparable, the former being dominated by the electronic width of the
narrow resonances. In principle this error could be further reduced by
the high luminosity measurements at BESS III. A further reduction of the
(already tiny) theory error, e.~g.~through a five-loop calculation
looks difficult. Further confidence in our result can be obtained from
the comparison with the forementioned lattice evaluation.

Improvements in the bottom quark mass determination could originate
from the experimental input, e.\ g.\ through an improved
determination of the electronic widths of the narrow $\Upsilon$
resonances or through a second, independent measurement of the $R$ ratio
in the region from the $\Upsilon(4S)$ up to 11.2 GeV. As shown in 
Fig.~\ref{fig:R_b},
there is a slight mismatch between the theory prediction above 11.2 GeV
and the data in the region below with their systematic error of less
than 3\%.

In this connection it may be useful to collect the most important pieces of
evidence supporting this remarkably small error. 
Part of the discussion is applicable to both
charm and bottom, part is specific to only one of them. 
In particular for charm, but to some extent also for bottom, the $\mu$-dependence
of the result increases for the higher moments, starting with $n=4$, and 
dominates the total error. We will therefore concentrate on the moments 
$n=1$, 2, and 3 which were used for the mass determination, 
results for $n=4$ will only be mentioned for illustration. 

\begin{figure}[ht]
\begin{center}
  \begin{tabular}{c}
    \leavevmode
    \epsfxsize=0.65\textwidth
    \epsffile{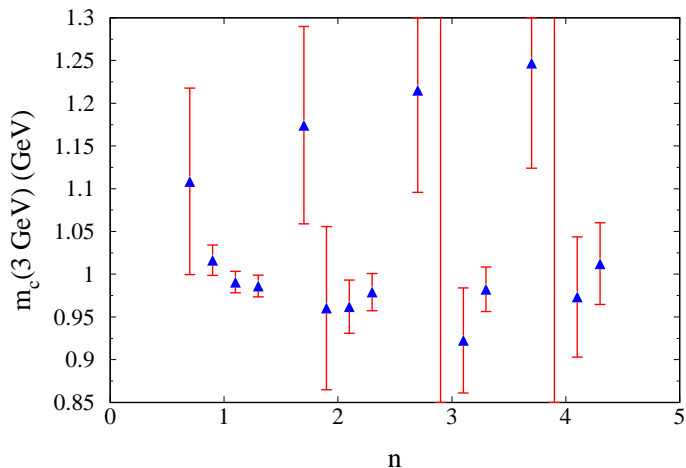}
  \end{tabular}
\end{center}
\vspace*{-2em}
\caption{
$m_c(3~\mbox{GeV})$ for $n=1,2,3$ and $4$.
For each value of $n$ the results from left to right correspond
the inclusion of terms of order $\alpha_s^0$, $\alpha_s^1$,
$\alpha_s^2$ and $\alpha_s^3$.
}
\label{fig::mom}
\end{figure}

Let us start with charm. Right at the beginning it should be
emphasized that the primary quantity to be determined is the 
running mass at the scale of 3~GeV, 
the scale characteristic for the production threshold and thus for
the process. Furthermore, at this scale the strong coupling 
$\alpha_s(3\;{\rm GeV}) =  0.258$ is already
sufficiently small such that the higher order terms in the perturbative series
decrease rapidly. Last not least, for many other observable of interest, like
$B$-meson decays into charm, or processes involving virtual charm quarks like
$B\to X_s\gamma$ or $K\to \pi \nu\bar \nu$, the characteristic scale is of order
3~GeV or higher. Artificially running the mass first down to 
${\cal O} (1\;{\rm GeV})$ and then back to a higher scale thus leads to an
unnecessary inflation of the error.

The quark mass determination is affected by the theory uncertainty, 
resulting in particular from our ignorance of yet uncalculated higher orders,
and by the error in the evaluation of the experimental moments. 
The former has been estimated \cite{Kuhn:2007vp}
by evaluating $m_c(\mu)$ at different renormalization scales
between 2 and 4~GeV (changing of course the coefficients $\bar{C}_n$
appropriately) and subsequently evolving $m_c(\mu)$ to $m_c(3\;{\rm GeV})$. The
error estimates based on these considerations are listed in Table~\ref{tab:mc1}.

The stability of the result upon inclusion of higher orders is also evident from 
Fig.~\ref{fig::mom} where the results from different values of $n$ 
are displayed separately in order $\alpha_s^i$ with $i=0$, 1, 2 and 3.
This argument can be made more quantitatively by rewriting eq.~(\ref{eq:m_Q}) 
in the form
\begin{eqnarray}
m_c &=& 
\frac{1}{2}
\bigg( \frac{9 Q_c^2}{4}\frac{C_n^{Born}}{{\cal M}_n^{\rm exp}}\bigg)^\frac{1}{2n} 
(1 + r^{(1)}_n \alpha_s + r^{(2)}_n\alpha_s^2 + r^{(3)}_n \alpha_s^3) 
\nonumber\\
&\propto&
1 - 
\left(\begin{array}{c}0.328\\0.524\\0.618\\0.662\end{array}\right)\alpha_s - 
\left(\begin{array}{c}0.306\\0.409\\0.510\\0.575\end{array}\right)\alpha_s^2 -
\left(\begin{array}{c}0.262\\0.230\\0.299\\0.396\end{array}\right)\alpha_s^3,
\label{mc} 
\end{eqnarray}
where the entries correspond to the moments with $n=1$, 2, 3 and 4. 
Note, that the coefficients are decreasing with increasing order of
$\alpha_s$. Estimating
the relative error through $r_n^{max}\alpha_s(3\;{\rm GeV})^4$ leads to
1.4 /  2.3 / 2.7 / 2.9  permille and thus to an estimate clearly
smaller than the one based on the $\mu$-dependence. 

\begin{figure}[t]
\begin{center}
  \begin{tabular}{c}
    \leavevmode
    \epsfxsize=.8\textwidth
    \epsffile{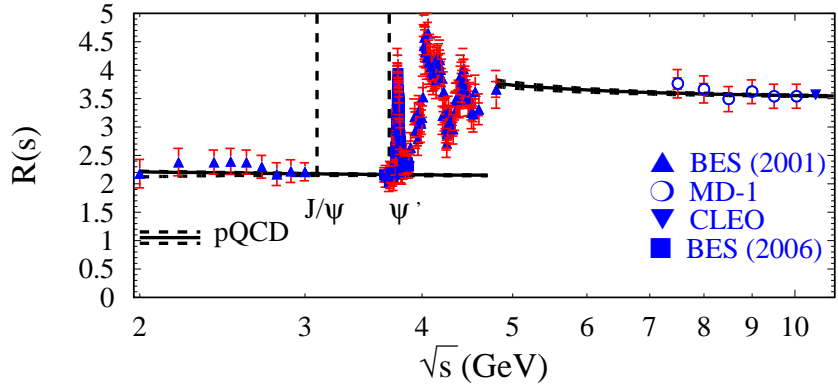}
    \\[-2em]
  \end{tabular}
\end{center}
\caption{$R(s)$ for different energy intervals around the charm threshold 
region. The solid line corresponds to the theoretical prediction.
}
\label{fig::R}
\end{figure}

The consistency between the results for different values of $n$ is another
piece of evidence (Fig.~\ref{fig::mom} and Table~\ref{tab:mc1}). 
For the lowest three moments the variation between the
maximal and the minimal value amounts to 10~MeV only. This, in addition,
points to the selfconsistency of our data set. Let us illustrate this
aspect by a critical discussion of the "continuum contribution",  i.e. the
region above 4.8 GeV, where data points are available at widely separated
points only. Instead of experimental data the theory prediction for $R(s)$
has been employed for the evaluation of the contribution to the moments.
If the true contribution from this region would be shifted down by, say,
10\%, this would move $m_c$, as derived from
$n=1$, up by about 20~MeV.  However, this same shift would lead to a
small increase by 3~MeV for  $n=2$ and leave the results 
for higher $n$ higher practically unchanged.
Furthermore, theory predictions and measurements in the region from
4.8~GeV up to the bottom-meson threshold, wherever available,
are in excellent agreement, as shown in Fig.~\ref{fig::R}, with deviations 
well within the statistical and systematical error of 3 to 5\%. 
Last not least, the result described above is in perfect agreement with
the recent lattice determination mentioned above.

Let us now discuss beauty, with $m_b$ evaluated at 
$\mu=10\;{\rm GeV}$. Again we first study the stability of the perturbative
expansion, subsequently the consistency of the experimental input.
With $\alpha_s(10\;{\rm GeV})=0.180$ the higher order corrections decrease
even more rapidly. 
Varying the scale $\mu$ between 5 and 15~GeV leads to a shift between 2 and 
6~MeV (Table~\ref{tab:mc1}) which is completely negligible. 
Alternatively we may consider the analogue of eq.~(\ref{mc}) with the correction 
factor
\begin{equation}
m_b/m_b^{\rm Born}=
1 - 
\left(\begin{array}{c}0.270\\0.456\\0.546\\0.603\end{array}\right)\alpha_s - 
\left(\begin{array}{c}0.206\\0.272\\0.348\\0.410\end{array}\right)\alpha_s^2 +
\left(\begin{array}{c}-0.064\\0.048\\0.051\\0.012\end{array}\right)\alpha_s^3.
\end{equation}
Taking $r_n^{max}\alpha_s^4$ for an error 
estimate leads to a relative error of .28 / .48 / .57 / 0.63 permille 
for $n=1$, 2, 3 and 4 respectively, which is again smaller
than our previous estimate. 
Let us now move to a
critical discussion of the experimental input. The contribution from the
lowest four $\Upsilon$-resonances has been taken directly from 
PDG~\cite{Yao:2006px} with systematic errors of the lowest three added 
linearly. 
The analysis~\cite{Chetyrkin:2009fv} of a recent 
measurement~\cite{:2008hx} of $R_b$ in the threshold region 
up to 11.20~GeV has provided results consistent with the earlier
analysis \cite{Kuhn:2007vp} but has lead to a significant reduction of 
the error in $m_b$.

In comparison with the charm analysis a larger
contribution arises from the region where data are substituted by the 
theoretically predicted $R_b$ with relative contributions of 
63, 41, 26  and 17 percent for $n=1$, 2, 3 and 4 respectively.  
This is particularly valid for the lowest moment.
For this reason we prefer to use the result from
$n=2$, alternatively we could have also taken the one from $n=3$. Let us
now collect the arguments in favour of this approach:

{\it i)} 
For light and charmed quarks the prediction for $R$ based on 
pQCD works extremely well already two to three GeV above threshold. 
No systematic shift has been observed between theory and experiment,
in the case of massless quarks, starting from around 2~GeV, and for the 
cross section including charm at and above 5~GeV up to the bottom 
threshold (Fig.~\ref{fig::R}). 
It is thus highly unplausible that the same approach should fail for 
bottom production.

{\it ii)} 
If the true $R_b$ in the continuum above 11.2~GeV would
differ from the theory prediction by a sizable amount,  the
results for $n=1$, 2 and 3 would be mutually inconsistent. Specifically,
a shift of the continuum term by 5\% would move $m_b$, derived from
$n=1$,  2 and 3 by about 64~MeV,  21~MeV and  9~MeV respectively.

{\it To summarize:}
Charm and bottom quark mass determinations have made significant progress
during the past years. A further reduction of the theoretical and experimental
error seems difficult at present. However, independent experimental results on
the $R$ ratio would help to further consolidate the present situation.
The confirmation 
by a recent  lattice analysis with similarly small uncertainty gives 
additional confidence in the result for $m_c$.

{\it Acknowledgments:} I would like to thank K.~Chetyrkin, A.~Maier,
P.~Maier\-h\"ofer, P.~Marquard, M.~Steinhauser and C.~Sturm for their 
collaboration and S.~Uccirati for help in preparing this manuscript.

\end{document}